\documentclass[twocolumn,showpacs,prl,aps,amsmath,superscriptaddress]{revtex4}

\usepackage{graphicx}% Include figure files
\usepackage{color,ulem}
\renewcommand{\Im} {\mathop{\mathrm{Im}}}
\newcommand{\cR}{\mathcal{R}}
\newcommand{\cV}{\mathcal{V}}
\newcommand{\cT}{\mathcal{T}}

\newcommand{\br}{\mathbf{r}}

\usepackage{bm}% bold math

\begin{document}

\title{Charge transport through weakly open one dimensional quantum
wires}
\author{N. B. Kopnin}
\affiliation{Low Temperature Laboratory,
Helsinki University of Technology, PO Box 5100, 02015 TKK, Finland}
\affiliation{ L. D.  Landau Institute for Theoretical Physics, 117940
Moscow, Russia}
\author{Y.  M. Galperin}
\affiliation{Department of Physics, University of Oslo, PO Box
1048 Blindern, 0316 Oslo, Norway and \\ A. F. Ioffe
Physico-Technical Institute of Russian Academy of Sciences, 194021
St. Petersburg, Russia}
\affiliation{Argonne National Laboratory,
9700 S. Cass Av., Argonne, IL 60439, USA}
\author{V. M. Vinokur}
\affiliation{Argonne National Laboratory, 9700
S. Cass Av., Argonne, IL 60439, USA}

\date{\today}

\begin{abstract} We consider resonant transmission through a
finite-length quantum wire connected to leads via finite transparency
junctions. The coherent electron transport is strongly modified by the
Coulomb interaction. The low-temperature current-voltage ($IV$) curves
show step-like dependence on the bias voltage determined by the
distance between the quantum levels inside the conductor, the pattern
being dependent on the ratio between the charging energy and level
spacing. If the system is tuned close to the resonance condition by
the gate voltage, the low-voltage $IV$ curve is Ohmic. At large
Coulomb energy and low temperatures, the conductance is
temperature-independent for any relationship between temperature,
level spacing, and coupling between the wire and the leads.
\end{abstract}

\maketitle

Quantization of conductance in ballistic one dimensional (1D)
channels or islands~\cite{disc} is one of the central issues of
current mesoscopic physics, see~\cite{reviews} for a review.  In
small-size islands, the Coulomb blockade suppresses conductivity
at low temperatures and small applied voltages~\cite{reviews2}.
The Coulomb blockade can be controlled by varying the charge of
the island between two tunnel barriers by the potential at the
gate electrode; the conductance of such system shows oscillations
as a function of the gate voltage, due to the periodic modulation
of the charging energy. The classical theory of the
Coulomb-blockade oscillations was developed in~\cite{Shekhter},
while the role of the discreteness of the spectrum was addressed
in \cite{GS,Beenakker91} and several subsequent papers. In these
models the island was assumed to be almost isolated from both
source and drain, so that the number of particles on it could be
considered a conserved quantity. This assumption, in general, does
not hold if the transparency of the contacts between the island
and the leads is finite~\cite{Matveev91}. Typical examples of such
systems are quantum dots~\cite{AGB}.

Transport through quantum conductors (QCs) essentially coupled to
the leads has not been fully investigated so far, though it offers
excellent opportunities for studying the interplay between quantum
and classical properties of QC. In this Letter, we investigate
transport through a relatively short single-mode QC assuming a
simplest model where charging effects are important whereas the
Luttinger-liquid behavior is still not pronounced (see,
e.g.,~\cite{LL}). We consider a ballistic conductor connected to
two leads via identical contacts with a finite transition
amplitude $\cT$. Its length $d$ is shorter that the electron mean
free path, the transport mechanism being the \textit{resonant
transmission}, which we distinguish from elastic cotunneling
\cite{AverinNazarov90}. This mechanism is relevant to recent
experiments on carbon nanotubes \cite{nanotubes1,nanotube}.

At $\cT \to 0$, the QC holds a fixed number of particles, $N$. If a
particle with energy $E$ tunnels between a lead and the state with
energy $\epsilon_p$ in the QC, the charging energy of the QC couples
$E$ and $\epsilon_p$ through \cite{Beenakker91}
\begin{equation} E=\epsilon_p +2E_C(N-N_0)\pm E_C
\label{E-above/below} %\, ,
\end{equation} where $\pm$ stands for adding/removing a particle,
$E_C=e^2/2C$ is the Coulomb energy of the QC, $C$ is the QC
capacitance, and $eN_0$ is the gate-induced charge density.  At finite
$\cal T$, the system forms a double-barrier resonant tunneling
structure, its properties being determined by the resonant states. The
distance between these resonant levels is $\Delta E=\pi \hbar v/d$
(where $v$ is the electron velocity) while their width is $\Gamma
=|{\cal T}|^2\Delta E$. The occupation probability of the modes is
determined by the competition between the relaxation processes in the
QC and in the leads, as well as by escape probability from the QC into
the leads. If the inverse escape time, $\Gamma/\hbar$, is much less
than the inter-mode relaxation time, $\tau^{-1}$, the occupation
probability is the equilibrium Fermi distribution with the lattice
temperature, the chemical potential being determined by the
(conserved) number of particles. In the opposite case where $\Gamma
\gg \hbar/\tau$ the distribution inside the QC is determined by the
coupling to the leads; it can be strongly nonequilibrium. The
distribution function is to be found from the kinetic equation. The
results, in general, depend on interplay between the different
relaxation mechanisms in the quantum conductor and the leads.

We will assume that the resonance width $\Gamma$ is small, but is
still larger than the relaxation rate in the QC,
\begin{equation}  \label{cond-distrib}
\hbar/\tau \ll \Gamma \ll
\Delta E \, .
\end{equation}
The requirement (\ref{cond-distrib}) is opposite to the condition
$\Gamma\ll \hbar/\tau $ considered in~\cite{Beenakker91}. In this
limit QC cannot be treated as an isolated quantum dot with the
vanishingly weak coupling to leads~\cite{end1}, and therefore, $N$
is not a good quantum number but rather an average value
determined by the interaction with the leads. To reflect this we
employ the Hartree-Fock type approach which leads to modification
of Eq.~(\ref{E-above/below}). We show that at finite $\cT$ the
excitation spectrum changes significantly. At large Coulomb
energy, $E_C>\Delta E$, and low temperatures, $T\ll E_C$, energy
exhibits a sharp step as a function of the internal momentum in
the QC. This step defines the internal Fermi level. The width of
the step is determined by the width of the resonant level. As a
result, zero-voltage conductance becomes temperature-independent
already at $T\ll E_C$ irrespective of the relationship between
$T$, $\Delta E$, and $\Gamma$. In a carbon nanotube placed between
the two leads the ratio $E_C/\Delta E= e^2d/2\pi \hbar v_F C$ can
reach values $e^2/2\pi \hbar v_F$ for the minimal capacitance of
the tube $C\sim d$. For typical $v_F =0.8\cdot 10^8$~cm/s
\cite{nanotube} this ratio is $\approx 0.46$, i.e., is of the
order of unity. Therefore, in order to achieve better
understanding of the experimental situation and having in mind
more general applications, both limits of large and small ratio
$E_C/\Delta E$ have to be studied.

\paragraph{Model.--}We specify the charging Hamiltonian as
\begin{equation} \hat H_C =E_{C} \left[ \sum _{\alpha} \int_\cV \hat
\psi ^\dagger_\alpha ({\bf r})\hat \psi _\alpha ({\bf r})\, d\cV
-N_0\right]^2 \label{chargeH}
\end{equation} $\cV$ is volume of the QC and $\alpha$ is the spin
index. The spin dependence is due to the level filling that controls
the charge on the QC. The spectral and transport properties are
determined by the retarded (advanced) and Keldysh Green functions
satisfying the equations
\begin{eqnarray*} &&\left\{ \begin{array}{ll} \delta ({\bf r}_1-{\bf
r}_2)\\0 \end{array}\right\}=G_0^{-1}(\epsilon,{\bf r}_1)\left\{
\begin{array}{ll} G_{\epsilon}^{R(A)}({\bf r}_1,{\bf r}_2)\\
G_{\epsilon}^{K}({\bf r}_1,{\bf r}_2)\end{array} \right\} \\ &&\quad
\quad + 2E_C\int d\cV \int \frac{d\epsilon_1}{4\pi i} G^K_{\epsilon_1}
({\bf r}_1,{\bf r}) \left\{ \begin{array}{ll} G_{\epsilon}^{R(A)}({\bf
r},{\bf r}_2)\\ G_{\epsilon}^{K}({\bf r},{\bf r}_2)\end{array}
\right\}\, , \\ && G_0^{-1}(\epsilon,{\bf r}_1) \equiv -\epsilon
+\hbar ^2 \hat{\mathbf{p}}_1^2/2m -\mu +U_C, \
\hat{\mathbf{p}}_1\equiv -i\partial/\partial \mathbf{r}_1.
\end{eqnarray*} The energy $U_C=e^2(N-N_0)/C$ is produced by the
charge on the QC, $\mu$ is the chemical potential; ${\bf r}_1$ belongs
to the QC, otherwise the charging energy vanishes.

\begin{figure}[t]
\centerline{
\includegraphics[width=0.53\columnwidth]{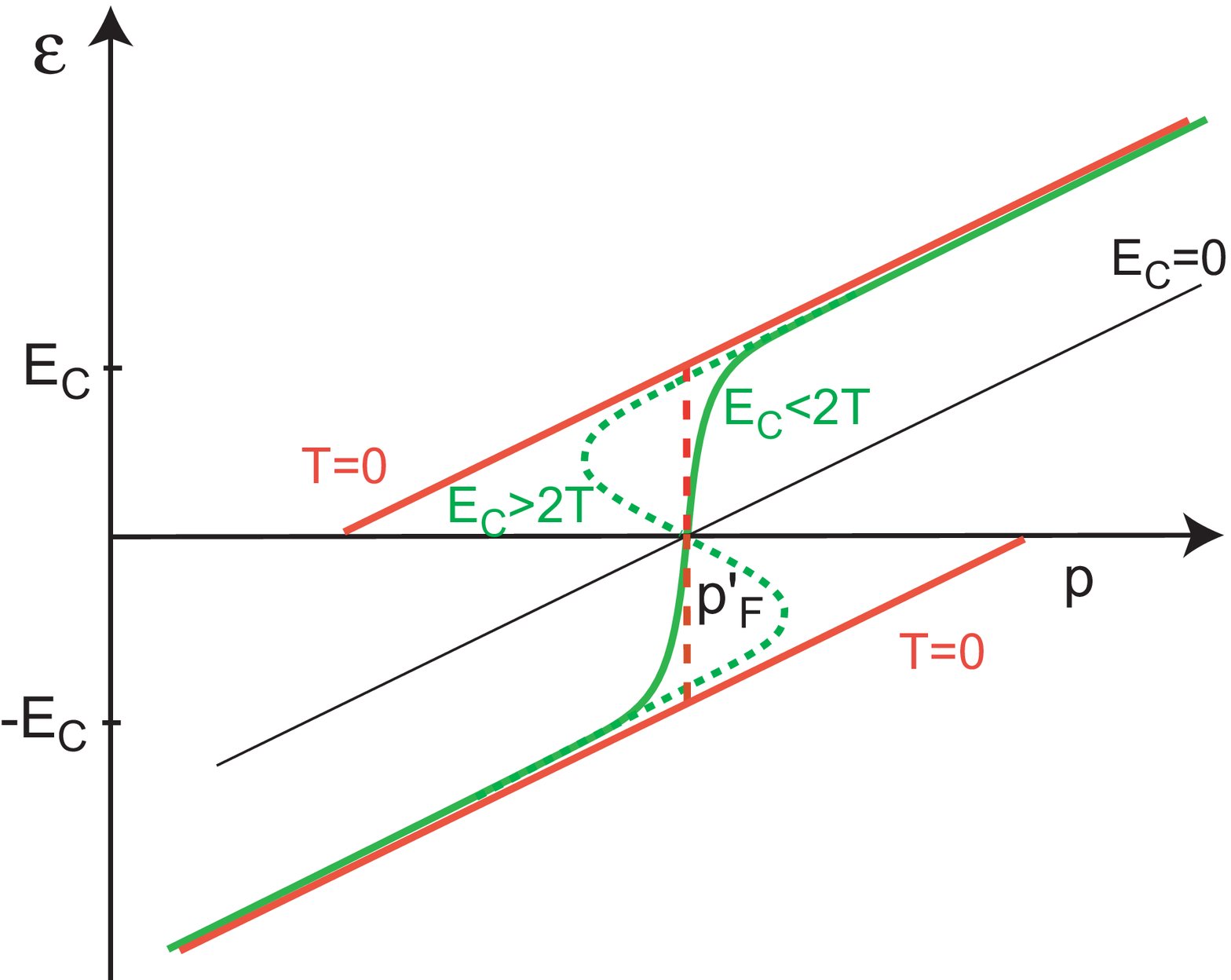}
\includegraphics[width=0.47\columnwidth]{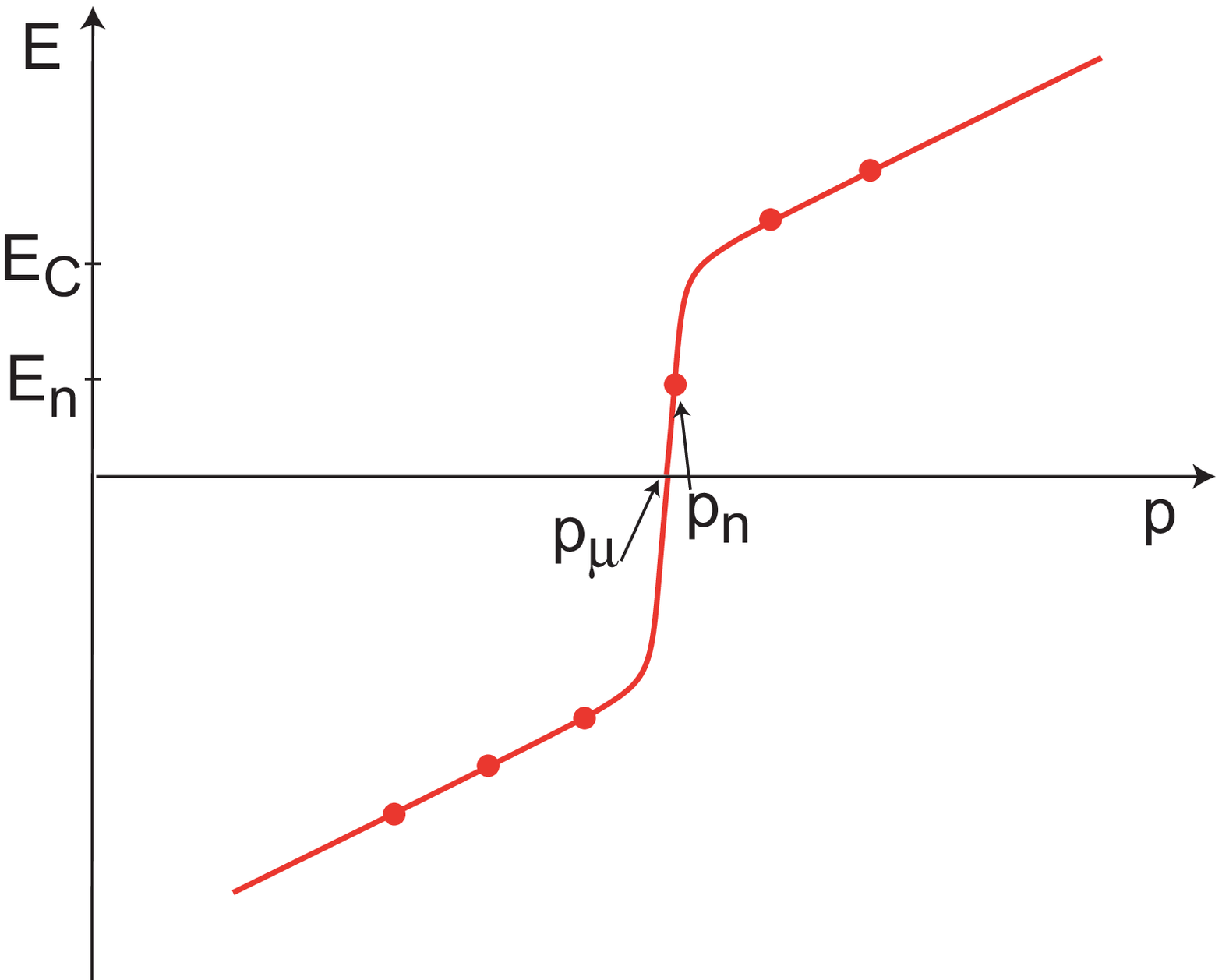}
} \caption{(Color online) (a)  Energy spectrum of an infinitely
long normal conductor. Finite temperature: The slope at $E=0$ is
positive for $E_C<2T$ (green line) and negative for $2T<E_C$
(dotted line). For $T=0$ (red lines) the Fermi surface corresponds
to the jump in the spectrum; $p_\mu ^\prime =p_\mu -U_C/v$ (dashed
vertical line). Straight (black) line: zero Coulomb energy. (b)
Energy spectrum for a finite-transparency contacts. The dots
illustrate sequential position of even and odd states.}
\label{fig-spectrum-F}
\end{figure}

For methodical purposes, let us start with the example of
infinitely long QC assuming $\Delta E \ll E_C$. Since $E_C$ is
coordinate independent one can consider the wave functions as
plane waves and use momentum representation. Then the Green
function has the pole at $\epsilon=\varepsilon_p$ where
\begin{equation} \varepsilon_p= \xi_p +U_C +E_C
\tanh\frac{\varepsilon_p}{2T}\  , \; \xi_p=\frac{\hbar ^2 p^2}{2m}
-E_F\ . \label{CoulGap1}
\end{equation} $\varepsilon_p$ is measured from the Fermi level
$\mu=E_F$ in the leads. At $\varepsilon_p \gg T$ this coincides with
Eq.~(\ref{E-above/below}). For $\varepsilon _p\rightarrow 0$ one has
$\varepsilon_p=(\xi_p +U_C)/(1-E_C/2T)$. The slope of the
$\varepsilon_p$ vs $\xi_p$ dependence is negative for $E_C>2T$, and
the function $\varepsilon_p$ becomes multi-valued, see
Fig.~\ref{fig-spectrum-F}(a). As we will see later, $\varepsilon_p$ in
fact experiences an abrupt jump as a function of internal momentum
shown in Fig.~\ref{fig-spectrum-F}(a) by the vertical line. This
infinite slope is a consequence of the chosen approximation of an
infinitely long QC with $\Delta E \to 0$.

In a finite QC, the jump acquires a finite width determined by the
transparency of the contacts. To study this problem in more detail it
is convenient to expand Green's functions over the orthonormal set of
functions $u_n(\br)$, which satisfy the Schr\"odinger equation
\begin{equation} \left(\frac{\hbar^2 \hat{\mathbf{p}}^2}{2m} -\mu
+U_C\right) u_n +E_C\sum_m u_m N_{mn} =E_n u_n\, . \label{BdG1}
\end{equation} Here $N_{mn}=\int d\cV \, u^*_m({\bf r} )u_n({\bf r} )
f(E_m)$; $f(E_m) \equiv 1-2n(E_m)$, where $n(E_m)$ is the occupation
probability of the $m$-th state. The diagonal element is proportional
to the average charge $eN_{nn}$ in the state $n$.

\paragraph{Spectrum.}-- Since main contributions to the transport come
from the quasi-resonant tunneling states it is convenient to expand
the functions $u_n$ over the scattering plane-wave states, $v_p(x)$,
satisfying the equation
\begin{equation} \left[\frac{\hbar^2\hat{\mathbf{p}}^2}{2m} -\mu +
V\sum_\pm \delta(x\pm d/2) \right] v_p(x) =\xi_p v_p(x)\, .
\label{sec-expansion}
\end{equation} The wave functions can be chosen as incident waves on
the left, $v_p$, and incident waves on the right, $v_{\bar{p}}$. We
assume a symmetric structure such that each barrier is characterized
by the same plane-wave reflection $\cR$ and transmission ${\cal T}$
amplitudes, $\cR=|\cR|e^{i\delta}$ and ${\cal T}=-|{\cal
T}|e^{i\delta}$ where $\delta$ is the scattering phase;
$|\cR|^2+|{\cal T}|^2=1$.
We now express the wave functions $u_n$ through the free functions
$v_p$: $u_n (x) =(2\pi)^{-1}\int A_{np} v_p (x)\,d p$. Using the
orthogonality and completeness of  both sets $u_n$ and $v_p$ one can
show that the states $u_n$ with the different $n$ are expanded in
$v_p$ with different $p$. This property, in particular, excludes the
multi-valued solutions of Eq. (\ref{CoulGap1}) shown in
Fig.~\ref{fig-spectrum-F}(a). Equation (\ref{BdG1}) can be then
rewritten in the form
\begin{eqnarray} \label{eq-expansion} &&\epsilon_p
-E_n=\int\frac{dp^\prime}{2\pi}\frac{A_{np^\prime}}{A_{np}}
\left[(e\varphi -U_C) M_{pp^\prime}- E_C Q_{pp^\prime}\right];
\nonumber \\ && M_{pp^\prime}\equiv\int_{-d/2}^{d/2}v_p^*(x)
v_{p^\prime} (x)\, dx\, , \label{exp-kernel} \\
&& Q_{pp^\prime}\equiv\sum _m
\int\frac{dp_1}{2\pi}\frac{dp_2}{2\pi} A_{mp_1}
A^{*}_{mp_2}f_1(E_m)M_{pp_1}M_{p_2p^\prime}. \nonumber
\end{eqnarray}
The matrix elements $M_{pp^\prime}$ can be explicitly
expressed through the transition amplitude, $\cT$.

In the equilibrium,  the distribution functions for particles
coming from the left and right are the same, $f(E)=\tanh (E/2T)$.
In what follows we concentrate on the situation when the
transparency satisfies Eq.~(\ref{cond-distrib}). For $|\cT |\ll 1$
there exist sharp resonances in the transmission when the momentum
of electrons inside the QC is close to the resonant values
corresponding to integer ratio $s\equiv (pd+\delta)/\pi$, $\delta
\approx \pi$. We denote $p_n^\pm$ the even and odd resonant states
for $s=2n+1$ and $s=2n$, respectively. At $E_C\gg \Delta E$ these
internal resonance momenta of $u_n(r)$ lie far from the incident
particle momenta in the leads $\sqrt{2m(E+E_F)}$ but are close to
the resonance momenta of the plane-wave states, $v_p(x)$. This
implies that transport is controlled only by discrete resonant
levels, which distinguishes the considered mechanism from the
usual elastic cotunneling through a multi-level quantum
dot~\cite{AverinNazarov90}. Consequently, one can put
$A_{np}\approx M_{q_n  p}^*=M_{p  q_n}$ where the continuous
parameter $q_n$ is the wave vector of a particle inside the QC
close to $n$-th resonance. Equation (\ref{eq-expansion}) for
$p=q_n$ becomes (for both even and odd states)
\begin{eqnarray}
&&E_n = \epsilon_{q_n} +U_C-e\varphi+ E_C \int
\frac{dp}{2\pi}f(E_p)M_{p p}^\pm\, ; \label{eq-expansion2} \\
&&M_{p p^\prime}^\pm = \frac{{\cal T}_{p^\prime}{\cal T}^*_{p}
 e^{i(p-p^\prime)d/2} \Delta(p-p^\prime)}{(e^{-ip^\prime d/2}\mp \cR e^{ip^\prime d/2})
(e^{ip d/2}\mp \cR^*e^{-ip
 d/2})}\, . \nonumber
\end{eqnarray}
Here $\epsilon_{q_n}= \hbar^2q_n^2/2m -E_F$, the integration is
performed over the vicinity of the $n$-th resonance, $p_n^\pm$;
$\Delta (p) \equiv (2/p)\sin (pd/2)$.
Equation~(\ref{eq-expansion2}) generalizes Eq.~(\ref{CoulGap1})
with the replacements $\varepsilon_p \to E_m$, $\xi_p \to
\epsilon_{q_n}$ and defines $q_n$ corresponding to the resonant
transmission through the QC.
%%%%%%%%%%%%%%%%%%%%%%%%%%%%%%%%
%%%%%%%%%%%%%%%%%%%%%%%%%%%%%%%%

If the  intra-QC relaxation is efficient, $\tau_\epsilon \ll
\hbar/\Gamma$, where $\tau_\epsilon$ is the inelastic relaxation
time in the QC, the distribution is determined by the states in
the QC, $f(E_p) \equiv f(\epsilon_{q_n})=\tanh
(\epsilon_{q_n}/2T)$.  The abrupt dependence shown in
Fig.~\ref{fig-spectrum-F}(a) is then smeared by the finite
temperature, and $E_n$ as a function of $q_n$ acquires a finite
width $\sim T$, similarly to the situation considered in
\cite{Beenakker91}. However, if Eq.~(\ref{cond-distrib}) holds,
and when temperature satisfies $T\ll E_C$, the spectrum $E_n(q_n)$
shows a jump over $2E_c$ at some value $q_n\equiv p_\mu$ such that
all the levels with $q_n>p_\mu$ are empty, while the levels with
$q_n<p_\mu$ are occupied. Therefore $f(E_n) \approx
\mathrm{sign}(q_n-p_\mu)$. With this approximation in the r.h.s.
of Eq.~(\ref{eq-expansion2}) one finds for $\cT \ll 1$
\begin{equation}
E_n = \hbar v(q_n-p_F)-e\varphi +U_C+E_C \Phi (p_n -p_\mu) \, .
\label{En-even/odd}
\end{equation}
Here $q_n$ is close to the resonance value $p_n\equiv \pi n/d$ and
\begin{equation}
\Phi(p_n -p_\mu)=(2/\pi)\arctan \left[2d(p_n-p_\mu)/|\cT|^2
\right]\ . \label{Phi}
\end{equation}
Equation (\ref{En-even/odd}) generalizes Eq.~(\ref{E-above/below})
for low transparency limited by Eq.~(\ref{cond-distrib});
Eqs.~(\ref{En-even/odd}) and (\ref{E-above/below}) coincide far
from the Fermi level. Energies $E_n$ for the resonance momenta
$p_n$ are shown in Fig.~\ref{fig-spectrum-F}(b). The width of this
function is determined by $\Gamma$, rather than by the
temperature.

The Fermi momentum,  $p_\mu$, is related to the number of
electrons inside the QC. We define the variation of the particle
number, $\delta N$, as a function of the bias and gate voltage.
Using Eqs.~(\ref{En-even/odd}), (\ref{Phi}) one can cast this
variation in the form $\delta N =-(1/2)\delta \left[\sum
_{n,\alpha} \Phi (p_n -p_\mu) \right]$. Here the sum runs also
over the electron spin index $\alpha$. Let us denote the average
number of electrons for zero gate voltage as $N_0$. If $N_0$ is an
integer, the chemical potential lies far from one of the
resonances $|\cT|^2/d\ll |p_\mu -p_n|\leq \pi n/2d $. For
non-integer $N_0$, one can split it into the integer part,
$[N_0]$, and the remainder, $\tilde{N}_0$. In this case, the
chemical potential is close to one of the resonances. Assuming
that the state $n$ is the closest to the Fermi level one can
determine $p_\mu$ from the equation $\Phi (p_n -p_\mu)=1-2\tilde
N_0$. In general,  the chemical potential and the number of
particles are related by the equation $2(N-[N_0])+\Phi (p_n
-p_\mu)=1$. Near the degeneracy point when $N-[N_0]=\pm 1/2$, the
Fermi momentum is close to one of the resonances.

\paragraph{Current.}For a finite bias $V$, energies in
the left and right leads are shifted by $\pm eV/2$. Following the
same approximation as above, we have instead of Eq.
(\ref{En-even/odd})
$$
E_n=\hbar v_F(q_n-p_F) -e\varphi +U_C +\frac{E_C }{2}\! \! \!
\sum_{i=L(R)}\! \! \! \Phi\left(p_n-p_\mu^{(i)}\right).
$$
Now the Fermi momenta, $p_\mu^L$ and $p_{\mu}^R$, are not equal
but are determined by the states in the left and right electrode,
respectively. Similarly to the equilibrium case, we get $\delta N
=-(1/4)\delta \left[\sum_{n,\alpha, i=L(R)}\Phi\left(p_n
-p_\mu^{(i)} \right)\right]$. Defining $j_p \equiv
u_p^*(x)[\partial u_p(x)/\partial x] $ one can express current as
\begin{equation}
I=-\frac{e\hbar}{m}\sum_{p,\alpha}\left[f_1^L(E_p) \Im j_p
+f_1^R(E_{\bar{p}}) \Im j_{\bar{p}} \right] , \label{def-current}
\end{equation}

\begin{figure}[t]
\centerline{\includegraphics[width=0.5\linewidth]{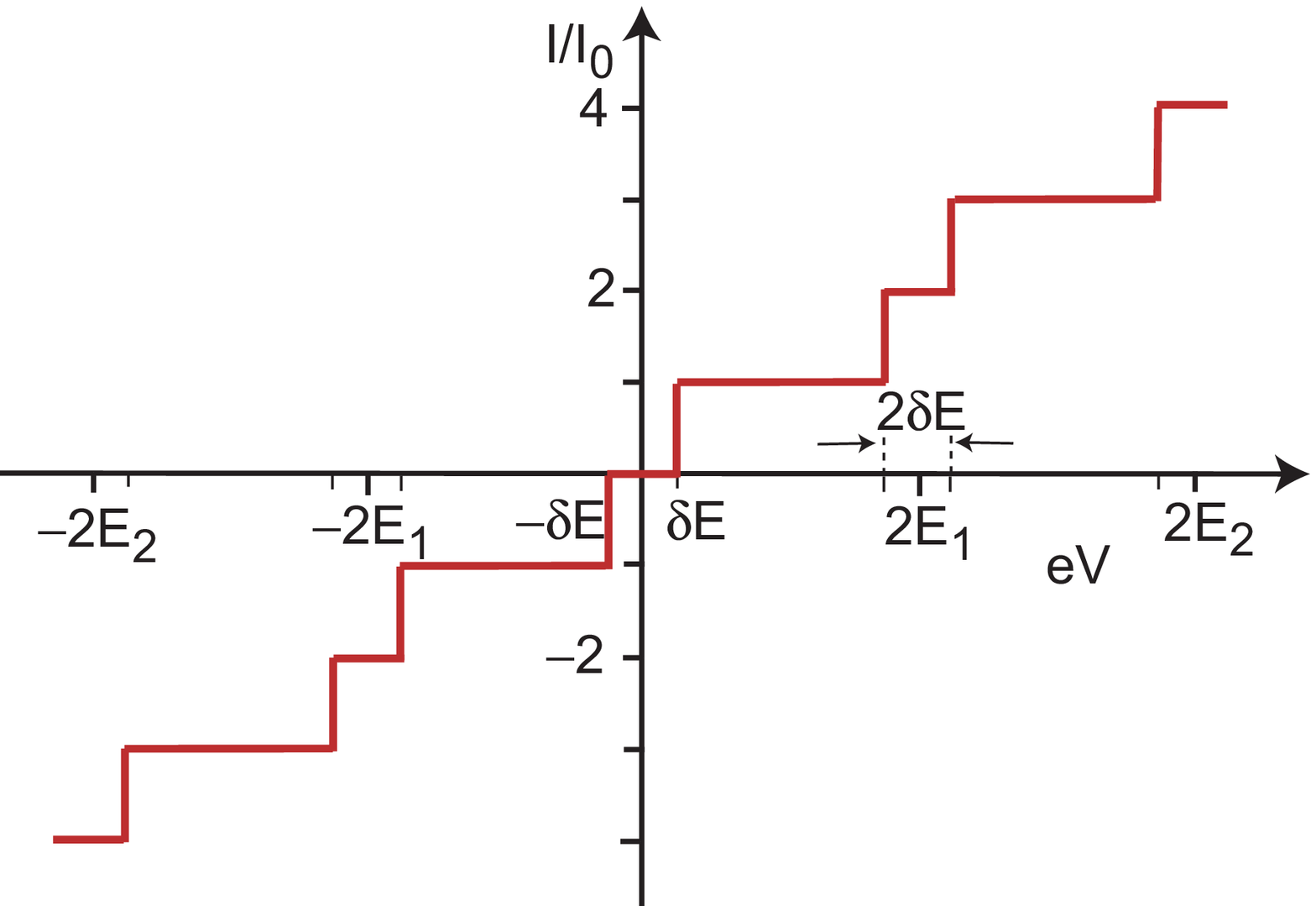}
\includegraphics[width=0.47\linewidth]{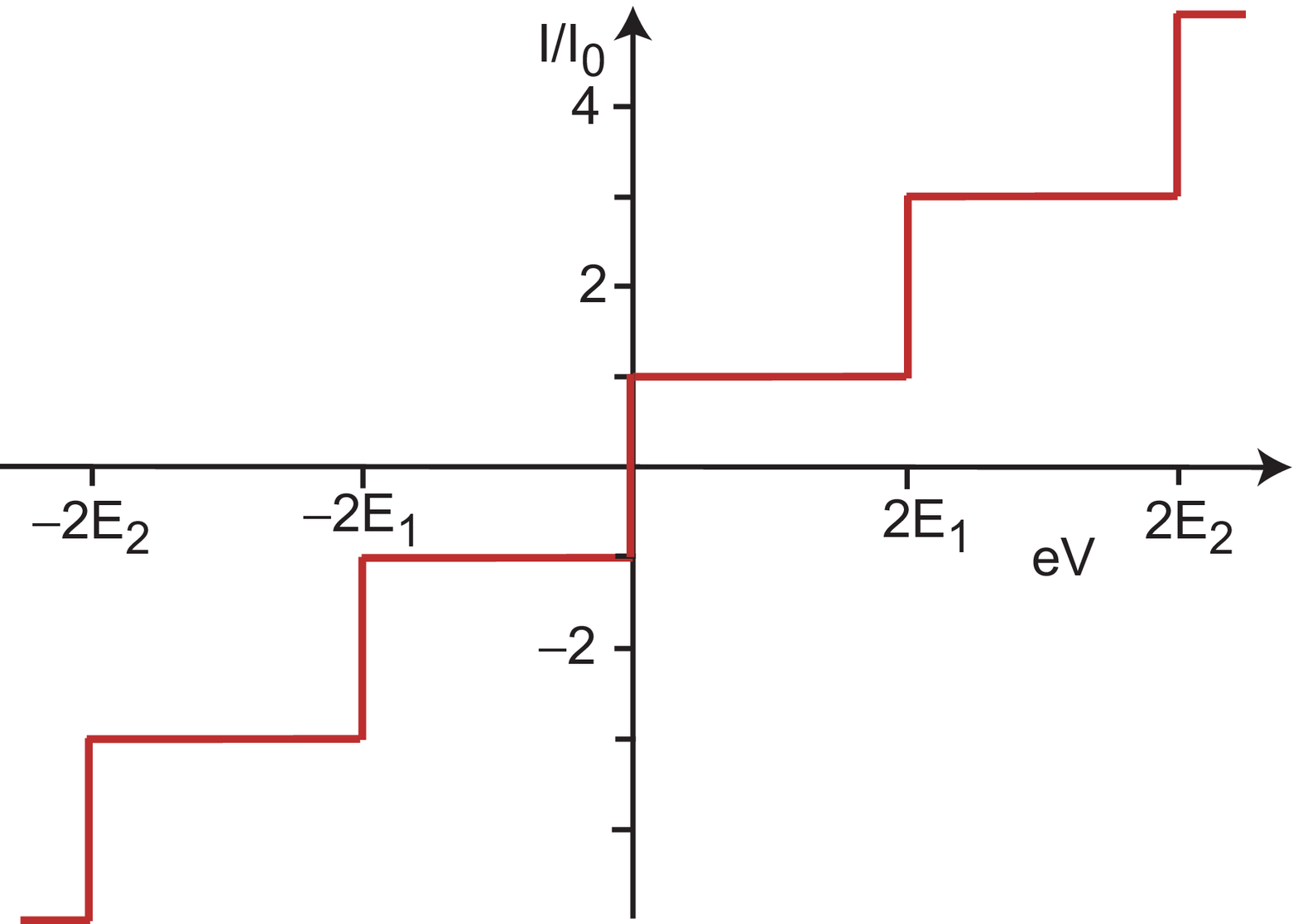}
} \caption{ (Color online) (a) Current steps for zero Coulomb
energy and $T=0$. $E_n = \Delta E n$ is the resonance energy, $
\delta E= \hbar v \alpha^\prime /d$ is the deviation from the
resonance. (b) For exact resonance, $\delta E=0$, the steps are
doubled.} \label{fig-steps1}
\end{figure}

The following analysis is different for small and large values of
$E_C/(\Delta E)$. At $E_C \ll \Delta E$, the energy $E_n =\hbar
v(q_n-p_\mu)$ is independent of the level population and thus of
the spin state. At relatively high temperatures,  $\Gamma \ll T\ll
\Delta E$, using Eq.~(\ref{def-current}), one obtains
\begin{equation}
I=
\frac{I_0}{2}
\sum _n\left[\tanh
  \frac{E_n^+}{2T} - \tanh
  \frac{E_n^-}{2T}\right], \ I_0\equiv \frac{e v}{2d}|{\cal T}|^2
.  \label{I-ballistic1}
\end{equation}
$E_n^\pm \equiv E_n \pm eV/2$, $n$ runs over even and odd states.
The $IV$-curves are shown in Fig.~\ref{fig-steps1}. The current
exhibits steps at $eV=2(\Delta E n \pm \delta E)$ where $\delta
E=(\hbar v/\pi)[p_\mu-\pi M/d]$ is the deviation from resonance
between the zero-voltage Fermi energy in the leads and one of the
levels, $M$ being an integer.
%%%%%%%%%%%%%%%%%%%%%%%%%%%%%%%%
%%%%%%%%%%%%%%%%%%%%%%%%%%%%%%%
For $T \ll \delta E$ and small voltages, $eV \le \delta E$, the
current is zero.  For the resonance, $\delta E=0$, the current is
given by the term with $n=M$ in the sum, $I= I_0 \tanh
\left(eV/4T\right)$. At $eV\ll T$ the conductance is Ohmic, $G=
eI_0/4T$, which agrees with Refs.~\cite{Beenakker91,CN2barrier}.

For large $E_C$, the states within the interval $\delta p\sim |\cT
|^2/d$ near the lowest resonance level, i.e., those which
contribute to the current, have energies $\sim E_C$. Therefore the
distribution function $\mathop{\mathrm{sign}}(q_n-p_\mu)$ can be
used up to temperatures $T\ll E_C$, regardless of the relation
between $T$ and $\Gamma$ or $\Delta E$. Using
Eq.~(\ref{def-current}) we find
\begin{equation}
I=(I_0/4)\sum_{n,\alpha} \left[ \Phi (p_n -p_\mu^R) - \Phi(p_n
-p_\mu^L) \right]\, . \label{current-fin1}
\end{equation}
The level positions are determined by the condition of minimal
total energy which fixes the value of $N-N_0$. Thus the levels, as
functions of the bias voltage, will cross the Fermi energy
pairwise, one level from above while the other from below, keeping
the number of electrons $N-N_0$ unchanged and lifting the Coulomb
blockade in the bias voltage. If the Fermi level lies between the
resonances, the first step of hight $I_0$ in the current appears
for $V=\Delta E/e$. The next steps appear when the bias voltage is
increased by $\delta V =2\Delta En/e$. The heights of the current
steps are the same as for low $E_C$: the extra factor
$\frac{1}{2}$ in Eq.(\ref{current-fin1}) is compensated by the
pairwise level crossing. However, the fine structure shown in
Fig.~\ref{fig-steps1}(a) disappears.

If the system is close to the degeneracy point $N-[N_0]=\pm 1/2$,
when one of the values $p_n$ is close to $p_\mu$, even a small
bias voltage is sufficient to produce current. We find from
Eq.~(\ref{current-fin1}) for low voltage, $eV\ll \Gamma $,
\begin{equation}
I= \frac{G_0V}{2} \frac{|{\cal T}|^4}{|{\cal T}|^4+4(p_n -p_\mu)^2
d^2} \, , \quad G_0 \equiv \frac{e^2}{\pi \hbar}\,
.\label{conduct-highC}
\end{equation}
In the resonance, $p_n =p_\mu$, the conductance is half of the
conductance quantum, $G_0/2$. The conductance does not depend on
temperature at $T \le E_C$. This fact differs from the
tunnel-approximation result of Ref.~\cite{Beenakker91} for the
temperature domain $\Gamma \ll T\ll E_C$. The difference is due to
the step in the energy spectrum caused by relatively strong
coupling to the leads. Probing the onset of Ohmic conductance at
the degeneracy point allows one to monitor the effective number of
electrons in the QC.

\paragraph{Conclusion.} We develop a mean field type
description of the Coulomb effects in the ``weakly open'' 1D
systems and analyze both the excitation energy spectrum and the
electric current.  The $IV$ curves show a step-like dependence on
the bias voltage, the exact shape being determined by the ratio
between the charging energy and the level spacing. At large
charging energy and low temperatures $T\ll E_C$, the low-voltage
Ohmic conductance, Eq.~(\ref{conduct-highC}), is
temperature-independent irrespectively to the magnitude of the
ratios $T/\Delta E$ and $T/\Gamma$.

\acknowledgments We thank K. A. Matveev and A. S. Mel'nikov for
helpful discussions. This work was supported by the ULTI program
under EU contract RITA-CT-2003-505313, by the U.S. Department of
Energy Office of Science contract No. DE-AC02-06CH11357, by the
Academy of Finland (grant 213496, Finnish Programme for Centers of
Excellence in Research 2002-2007/2006-2011), and by the Russian
Foundation for Basic Research grant 06-02-16002.

\end{document}